\documentstyle[]{mn}

\title[The stellar mass ratio of GK~Persei]{The stellar mass ratio of 
GK~Persei}

\author[L.\, Morales-Rueda et al.] {L.\, Morales-Rueda$^{1}$, M.\,D.\,
  Still$^{2,3,4}$, P.\, Roche$^{5}$,
  J.\,H.\, Wood$^{6,7}$, J.\,J.\, Lockley$^{1}$\\
  $^{1}$Dept of Physics and Astronomy, University of Southampton, UK 
  (lmr@astro.soton.ac.uk)\\
  $^{2}$NASA Goddard Space Flight Center, Code 662, Greenbelt, MD
  20771, USA
  (still@milkyway.gsfc.nasa.gov)\\
  $^{3}$Universities Space Research Association, 7501 Forbes Blvd,
  Suite 206, Seabrook, MD 20706, USA\\
  $^{4}$Physics and Astronomy, University of St Andrews, North Haugh,
  St Andrews, UK\\
  $^{5}$Dept. of Physics and Astronomy, University of Leicester, UK\\
  $^{6}$School of Chemistry and Physics, Keele University, UK\\
  $^{7}$Department of Astronomy, San Diego State University, San Diego, CA 92182, USA}

\date{Accepted 0000 000 00; Received 0000 000 00; in original
  form 0000 000 00}

\pagerange{\pageref{firstpage}--\pageref{lastpage}}

\def\LaTeX{L\kern-.36em\raise.3ex\hbox{a}\kern-.15em
    T\kern-.1667em\lower.7ex\hbox{E}\kern-.125emX}

\begin{document}

\newcommand{\gk}{\mbox{GK~Per}} 
\newcommand{\etal}{\mbox{et\ al.}}
\newcommand{\kmsec}{\,\mbox{$\mbox{km}\,\mbox{s}^{-1}$}}
\newcommand{\phispin}{$\phi_{\tiny spin}$}
\newcommand{\phiorb}{$\phi_{\tiny orb}$}
\newcommand{\hb}{\hbox{$\hbox{H}\beta$}}
\newcommand{\hgam}{\hbox{$\hbox{H}\gamma$}}
\newcommand{\heii}{\hbox{$\hbox{He\,{\sc ii}\,$\lambda$4686\,\AA}$}}
\newcommand{\hei}{\hbox{$\hbox{He\,{\sc i}\,$\lambda$4472\,\AA}$}}
\newcommand{\ciiiniii}{\hbox{$\hbox{C\,{\sc iii}/N\,{\sc
        iii}\,$\lambda\lambda$4640--50\,\AA}$}}

\label{firstpage}

\maketitle

\begin{abstract}
  We study the absorption lines present in the spectra of the
  long-period cataclysmic variable \gk\ during its quiescent state,
  which are associated with the secondary star.  By comparing
  quiescent data with outburst spectra we infer that the donor star
  appears identical during the two states and the inner face of the
  secondary star is not noticeably irradiated by flux from the
  accreting regions.  We obtain new values for the radial velocity
  semi-amplitude of the secondary star, K$_{\rm K}$ = 120.5 $\pm$ 0.7
  \kmsec, a projected rotational velocity, V$_{\rm K}\sin i$ = 61.5
  $\pm$ 11.8 \kmsec\ and consequently a measurement of the stellar
  mass ratio of \gk, $q = M_{\rm K} / M_{\rm WD}$ = 0.55 $\pm$ 0.21.
  The inferred white dwarf radial velocities are greater than those
  measured traditionally using the wings of Doppler-broadened emission
  lines suspected to originate in an accretion disk, highlighting the
  unsuitability of emission lines for mass determinations in
  cataclysmic variables.  We determine mass limits for both components
  in the binary, $\rm M_{\rm K} \geq 0.48 \pm 0.32 \rm M_{\odot}$ and
  $\rm M_{\rm WD} \geq 0.87 \pm 0.24 \rm M_{\odot}$.
  
\end{abstract}


\section{Introduction}

\gk\ (Nova Per 1901; Campbell 1903), is a cataclysmic variable (CV)
that undergoes dwarf nova outbursts every 2--3\,years, brightening
from 13th to 10th magnitude (Sabbadin \& Bianchini 1983).  Its
spectrum is a composite of broad emission lines and narrow absorption
lines that move sinusoidally in wavelength over the orbital cycle, but
180 degrees out of phase, indicating origins associated with both
stars.  The absorption lines are generally photospheric and from the
secondary star (Kraft, 1964), while the emission lines are probably
from an accretion disc around the compact object (we explore this
further in a future paper).  Within the sample of known CVs, \gk\ has
the longest orbital period, P$_{\mbox{orb}}$ = 1.996803\,d(0.000007),
(Crampton, Cowley \& Fisher 1986).  This makes observing the orbital
cycle from the ground difficult unless observations are separated over
several observing seasons or a number of observatories are employed
contemporaneously at varying longitudes.

Previously, (Morales-Rueda, Still \& Roche 1996, 1999; hereafter MSR96
and MSR99), we have presented spectrophotometric observations of \gk\ 
taken during its 1996 outburst (Mattei \etal 1996).  In the current
paper we present spectroscopic observations of \gk\ during its
quiescence state and focus mainly on the behaviour of the secondary
star absorption lines, searching for signs of heating on the inner
surface of the companion.  Furthermore we measure the radial velocity
of the secondary star and combine this with previously published
velocities. With all the data, covering from 1964 to 1995, we
calculate a new ephemeris for \gk\ and a more accurate value for the
radial velocity semi-amplitude of the companion star, K$_{\rm K}$.  We
measure V$_{\rm K}\sin i$ and consequently the mass ratio $q$ for \gk.
A lower limit for the white dwarf mass is derived from these values
which is significantly different to the mass inferred by emission line
measurements.  This provides a further example of the general
unsuitability of emission line for measuring stellar masses in CVs
(Dhillon, Marsh \& Jones, 1991).

\section{Observations}

Spectra of \gk\ during its quiescent state were acquired in 1995, from
October 11 to 13, and from October 31 to November 2.  The first set of
spectra was taken at the 2.5\,m Isaac Newton Telescope (INT) on La
Palma using the Intermediate Dispersion Spectrograph. The standard
readout mode was used with the Tektronix CCD TEK4 windowed to
1024\,$\times$\,340 pixels. Exposure times were 35\,s with 90\,s dead
time, and the spectral resolution at \hb\ was 110 \kmsec. The R1200B
grating with 1200\,lines\,mm$^{-1}$ covered the wavelength range from
$\lambda\lambda$4136 -- $\lambda$5000 \AA.  351 spectra of \gk\ and 8
spectra of K-type template stars were obtained during this observing
run. Several bias and flatfield frames were obtained each night to
correct for the bias level and the pixel-to-pixel response variations
of the chip, and a CuAr frame was taken every 15 frames of the target
to calibrate the spectra in wavelength.  A wide slit exposure of the
standard HD19445 (Oke \& Gunn, 1983) was taken every night to
calibrate the detector's larger scale response variations.  A
spectrograph slit orientation of PA~$249.1^{\circ}$ allowed a 15th
magnitude nearby star approximately 25\,arcsec ENE of \gk\ to be
employed as calibration for light losses on the slit. A wide slit
spectrum of \gk\ and the comparison star was also taken to calibrate
the absolute flux scale. After debiasing and flat-fielding the frames,
spectral extraction proceeded according to the optimal algorithm of
Horne (1986).

The second set of spectra was taken using the low-to-moderate
resolution spectrograph ES2 mounted on the 2.1\,m telescope at
McDonald Observatory in Texas. The TI1 CCD camera and a grating ruled
at 1200~lines\,mm$^{-1}$ were used, covering wavelengths
$\lambda\lambda$4196 -- $\lambda$4894 \AA. Exposure times were 350\,s
with 40\,s readout time, and the resolution at \hb\ was 130~\kmsec.
Bias and flatfield frames were taken each night with regularly-spaced
arc lamp exposures. The standard Feige~110 (Stone, 1977) was observed
to flux calibrate the data.  Simultaneous photometry of \gk\ was taken
during the campaign at the McDonald 36-inch telescope with the
Stiening Photometer (Zhang \etal, 1991) using UBVR filters (see
Skidmore \etal\ 1997, for a description of the filters).  The exposure
times for the photometry were 1\,s. The combination of the
spectroscopy and the photometry yielded 113 spectra of \gk\ calibrated
in flux.  Spectra of 7 K-type template stars were also obtained.
Tables~1 and 2 give a journal of the \gk\ observations and a list of
the K-type templates observed during both campaigns and their spectral
classes respectively.

\begin{table}
\caption{Journal of observations. The orbital phases were calculated
  using the ephemeris obtained in Section 3.2. Phase 0
corresponds to superior conjunction of the white dwarf. Labels (1) and
(2) indicate  which observation campaign the data were taken at: (1)
INT, (2) McDonald.}
\begin{center}
\begin{tabular}{rccccr}
\multicolumn{1}{c}{Date} & \multicolumn{1}{c}{Start} & 
\multicolumn{1}{c}{End} & \multicolumn{1}{c}{Start} & 
\multicolumn{1}{c}{End} & \multicolumn{1}{c}{No. of} \\
\multicolumn{1}{c}{(1995)} & \multicolumn{2}{c}{(UT)} & 
\multicolumn{2}{c}{(Orbital phase)} & \multicolumn{1}{c}{spectra} \\ \hline
11 Oct (1)&22.27 &4.58 &0.027 &0.158 &129 \\
12 Oct (1)&22.91 &6.48 &0.541 &0.699 &165 \\
13 Oct (1)& 3.62 &6.59 &0.140 &0.202 &57 \\
31 Oct (2)&5.76 &9.67  &0.199 &0.281 &30 \\
1  Nov (2)&5.22 &11.71 &0.689 &0.824 &51 \\
2  Nov (2)&4.88 &8.96  &0.183 &0.268 &32 \\
\end{tabular}
\end{center}
\end{table}

\begin{table}
\caption{K star templates observed, 
with their corresponding spectral classification. (a) Houk \&
Smith-Moore (1988); (b) Nassau \& van Albada (1947); (c) Yoss (1961);
(d) Roman (1952); (e) Georgelin (1967), (f) Griffin \& Redman (1960),
(g) Hoffleit \& Warren (1991).}
\begin{center}
\begin{tabular}{llllll}
\multicolumn{1}{l}{Name} & \multicolumn{1}{l}{SpType} &
\multicolumn{1}{l}{Ref.} &
\multicolumn{1}{l}{Name} & \multicolumn{1}{l}{SpType} &
\multicolumn{1}{l}{Ref.}\\
\hline
\multicolumn{3}{c}{October 1995} & \multicolumn{3}{c}{November 1995} \\
\hline
HR190  & K1{\sc iii} & a   & 13 Lac     & K0{\sc iii}   & d, f\\
HR8688 & K1{\sc iii} & b   & 1 Peg      & K1{\sc iii}   & d, f\\
HR8415 & K2{\sc iii} & c   & 69 Aql     & K2{\sc iii}   & d \\
HR8632 & K3{\sc iii} & d   & 39 Cyg     & K3{\sc iii}   & d \\
HR8974 & K1{\sc iv}  & d   & 33 Vul     & K3.5{\sc iii} & g \\
HR8881 & K1{\sc v}   & e   & 3$\eta$Cep & K0{\sc iv}    & d, f\\
HR222  & K2{\sc v}   & f   & HD197964   & K1{\sc iv}    & g \\
HR8832 & K3{\sc v}   & d, f&            &               &  \\
\end{tabular}
\end{center}
\end{table}

\section{Results}

\subsection{Spectral classification of the donor star}
\label{donor:speclass}

\begin{figure*}
\begin{picture}(100,0)(10,20)
\put(0,0){\includegraphics{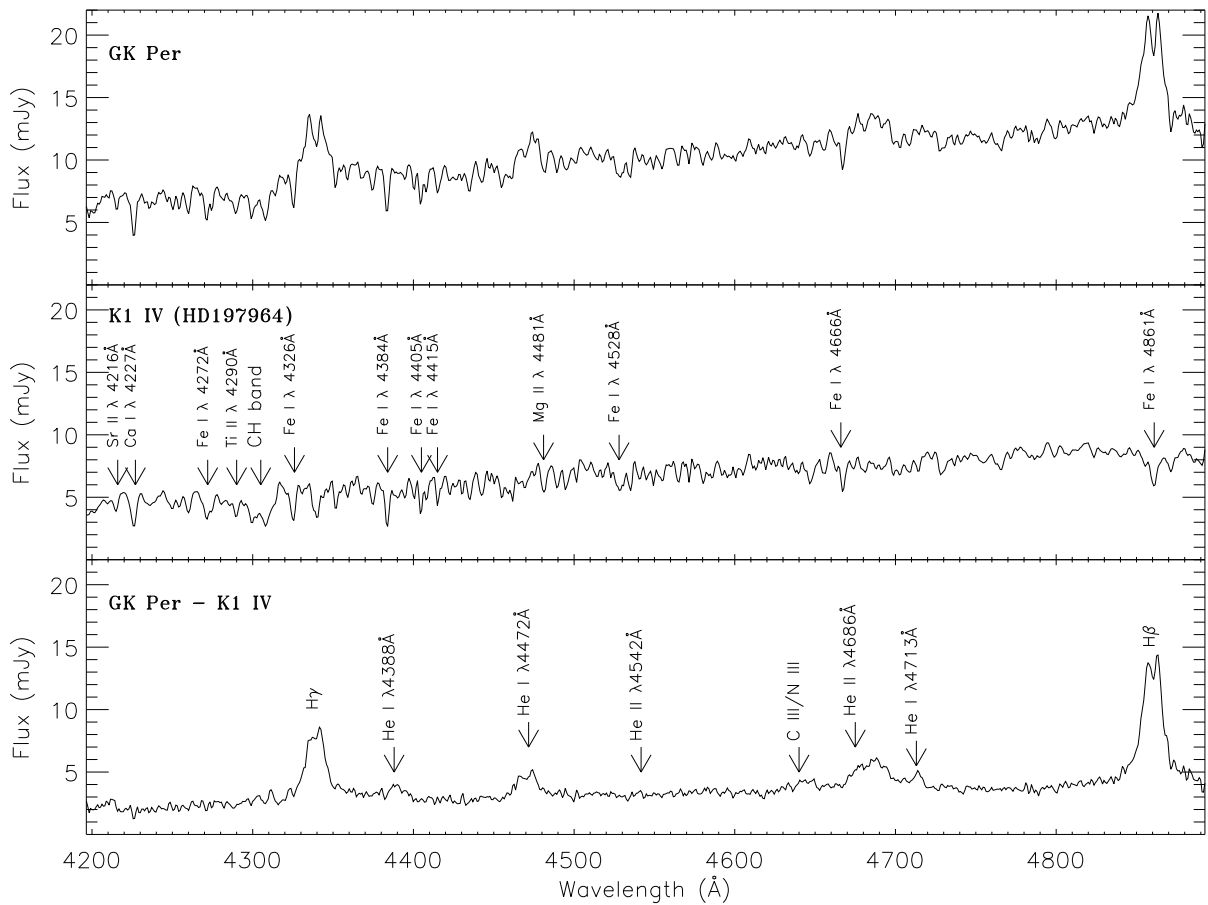}}
\noindent
\end{picture}
\vspace{125mm}
\caption{The top spectrum is an average of the spectra of \gk\ 
  obtained during the first night of observations on the October 1995
  campaign at the INT. The middle spectrum corresponds to the K1{\sc
    iv} template HD197964 multiplied by a constant, 0.3$\times
  10^{-3}$.  The bottom spectrum is the residual resulting from the
  subtraction of the template from the averaged \gk\ spectrum and
  probably resembles the spectrum of the accretion flow during
  quiescence. Some absorption and emission lines present in the
  spectra have been labelled. Most of the absorption lines were
  identified by Crampton et al. (1986).}
\label{donor:speclass:quiescence:fig1}
\end{figure*}

In 1964 Kraft noticed the presence of absorption features
characteristic of K-type stars in the optical spectrum of \gk.  Kraft
(1964), Gallagher \& Oinas (1974), Crampton \etal\ (1986) and Reinsch
(1994) carried out temporal studies of these absorption lines and
concluded that their origin was the mass-donating secondary star of
the binary system.  These authors classified the donor star by
comparing its absorption features with the same features seen in
K-type template stars.  Kraft (1964), Gallagher \& Oinas (1974) and
Reinsch (1994) obtained consistent classifications; K2{\sc iv}p,
K2{\sc iv}p - K2{\sc v}, and K2{\sc v} or K3{\sc v} respectively.
Kraft (1964) and Gallagher \& Oinas (1974) noted changes in the line
ratios from spectrum to spectrum throughout their observing campaign
and the classification they gave was the mean spectral type of all the
observations. The spectral type of the secondary as measured by
Crampton \etal\ (1986) (K0{\sc iii} - K0{\sc iv}) is not consistent
with those of the authors mentioned above.  Moreover, Crampton \etal
(1986) conclude that phase-related spectral changes seen by previous
authors are not present in their own data.

The spectra of \gk\ and the K-type templates presented here were
binned into constant velocity bins.  We used the fit to the orbital
radial velocity of the secondary calculated in Section 3.2 (systemic
velocity $\gamma$=40.8 \kmsec, radial velocity semi-amplitude of the
companion K$_{\rm K}$=120.5 \kmsec) to shift out the orbital motion of
the absorption lines with a quadratic rebinning algorithm.  The \gk\ 
spectra from each night were then binned into two orbital phase
intervals to investigate the possible changes in spectral and
luminosity classes of the secondary at different orbital phases. We
selected the regions where there were mostly-uncontaminated absorption
features, i.e.  regions containing obvious emission lines
($\lambda\lambda$4265--4399\AA, $\lambda\lambda$4435--4484\AA,
$\lambda\lambda$4503--4553\AA, $\lambda\lambda$4620--4727\AA,
$\lambda\lambda$4842--4878\AA, and $\lambda\lambda$4910--4935\AA) were
masked out from the data.

\begin{figure*}
\begin{picture}(100,0)(10,20)
\put(0,0){\includegraphics{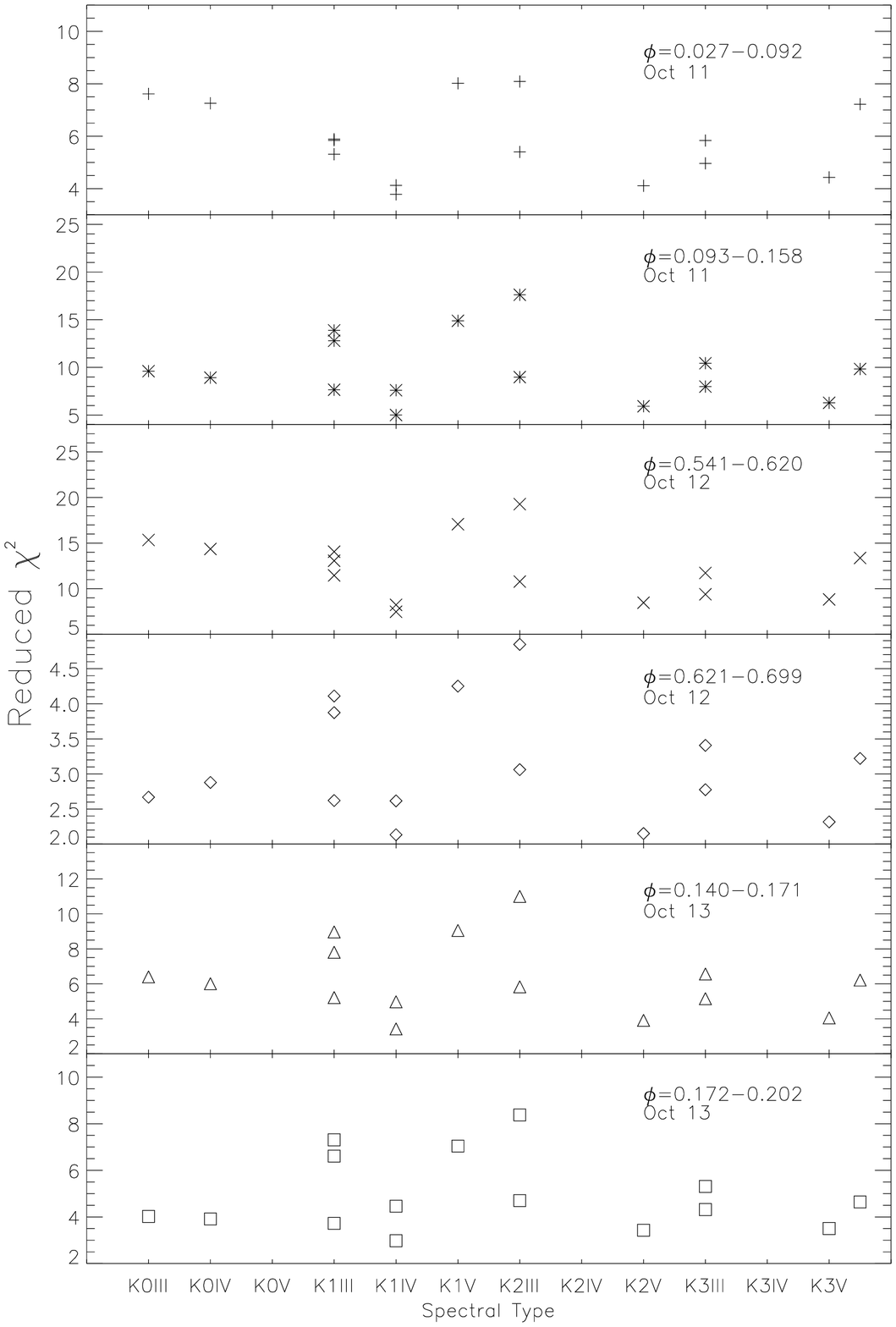}}
\put(0,0){\includegraphics{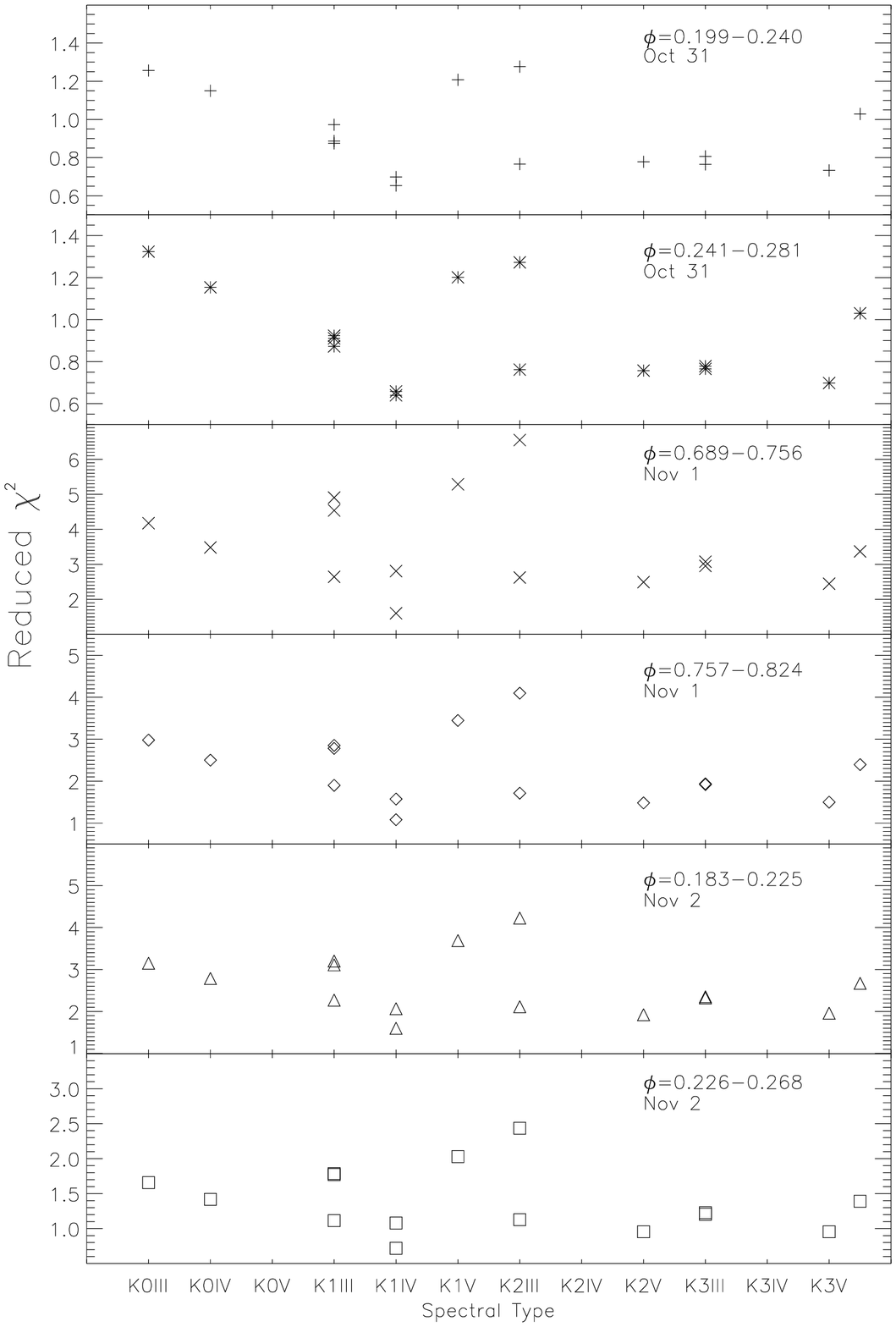}}
\noindent
\end{picture}
\vspace{130mm}
\caption{$\chi^2$ per degree of freedom for each template, each night 
  after applying optimal subtraction to the quiescent data. Note that
  the $\chi^2$ for the two orbital bins for each night is given.
  $\phi$ is the orbital phase range of each bin.}
\label{donor:speclass:quiescence:chi}
\end{figure*}

In order to classify the secondary star, an optimal subtraction method
(Marsh, Robinson \& Wood 1994) was applied.  The K star templates are
individually subtracted from the total spectrum.  The residual is
compared to a smoothed version of itself, produced by convolving the
residual with a Gaussian of FWHM = 13\,\AA\ using the $\chi^2$
statistic.  For each K star, this procedure is iterative, where the
template is scaled by a multiplicative free parameter which is
determined by minimising $\chi^2$.  The assumption is that the
correctly-scaled template spectrum is a good representation of the
secondary star and the most intrinsically-smooth residual is a good
representation of the accretion flow spectrum.  The template providing
the smallest minimum-$\chi^2$ is considered the best match to the
secondary star spectrum.  A plot of the minimum-$\chi^2$ per degree of
freedom found for each template for the two orbital phase intervals
observed each night can be found in
Fig.~\ref{donor:speclass:quiescence:chi}.

The best-fit template is a K1{\sc iv} star (HD197964) in all cases.
The companion star contributes 69 percent of the total light at
4550\AA, in contrast with 13 percent measured during outburst (MSR99).
Fig.~1 presents the total spectrum of \gk\ before (top panel) and
after (bottom panel) the subtraction of the scaled K1{\sc iv} star
template (middle panel).  Identified emission and absorption lines are
labelled.  This spectral classification of the secondary is close to
previous studies of \gk\ during quiescence.  This result does not fit
within the spectral class-orbital period relation fit by Smith \&
Dhillon (1998). This may indicate that the secondary star in \gk\ is
not a main-sequence star, which is not surprising as basic
evolutionary theories tell us that one should not find a main-sequence
star in a CV with an orbital period much longer than $\sim$9 hours
(Patterson 1984).

The INT data covers approximately the same orbital phases as the 1996
outburst data analysed in MSR96 and MSR99.  The template star nearest
in spectral class to the donor during outburst at phases, $\phi \sim$
0.14--0.18 and $\phi \sim$ 0.64--0.68 is also a K1{\sc iv} star
(MSR99).  The secondary star shows the same spectral class during both
outburst and quiescence at those particular phases.  Furthermore,
there is no convincing evidence for the rapid spectral changes
reported by Kraft (1964) and Gallagher \& Oinas (1974).

\subsection{Radial velocity of the secondary star}
\label{donor:rv:quiescence}

The continuum was subtracted from individual spectra by masking
wavelength regions with major emission lines, fitting a spline with 3
knots and subtracting the fits from the \gk\ spectra. For the
template, a two-knot spline fit was subtracted.  The spectra were
binned in velocity to ensure that all covered the same wavelength
region and had the same logarithmic dispersion. Emission line regions
remained masked out and the absorption lines of \gk\ were
cross-correlated (Tonry \& Davis 1979) against those of the K1{\sc iv}
template. The template spectrum was broadened by the value of V$_{\rm
  K}\sin$ i obtained in Section.~3.4 before cross-correlation was
carried out.  This is an iterative process because to measure V$_{\rm
  K}\sin$ i we need to know K$_{\rm K}$. We carried out three
iterations after which the values of K$_{\rm K}$ and V$_{\rm K}\sin$ i
were stable.

\begin{figure*}
\begin{picture}(100,0)(10,20)
\put(0,0){\includegraphics{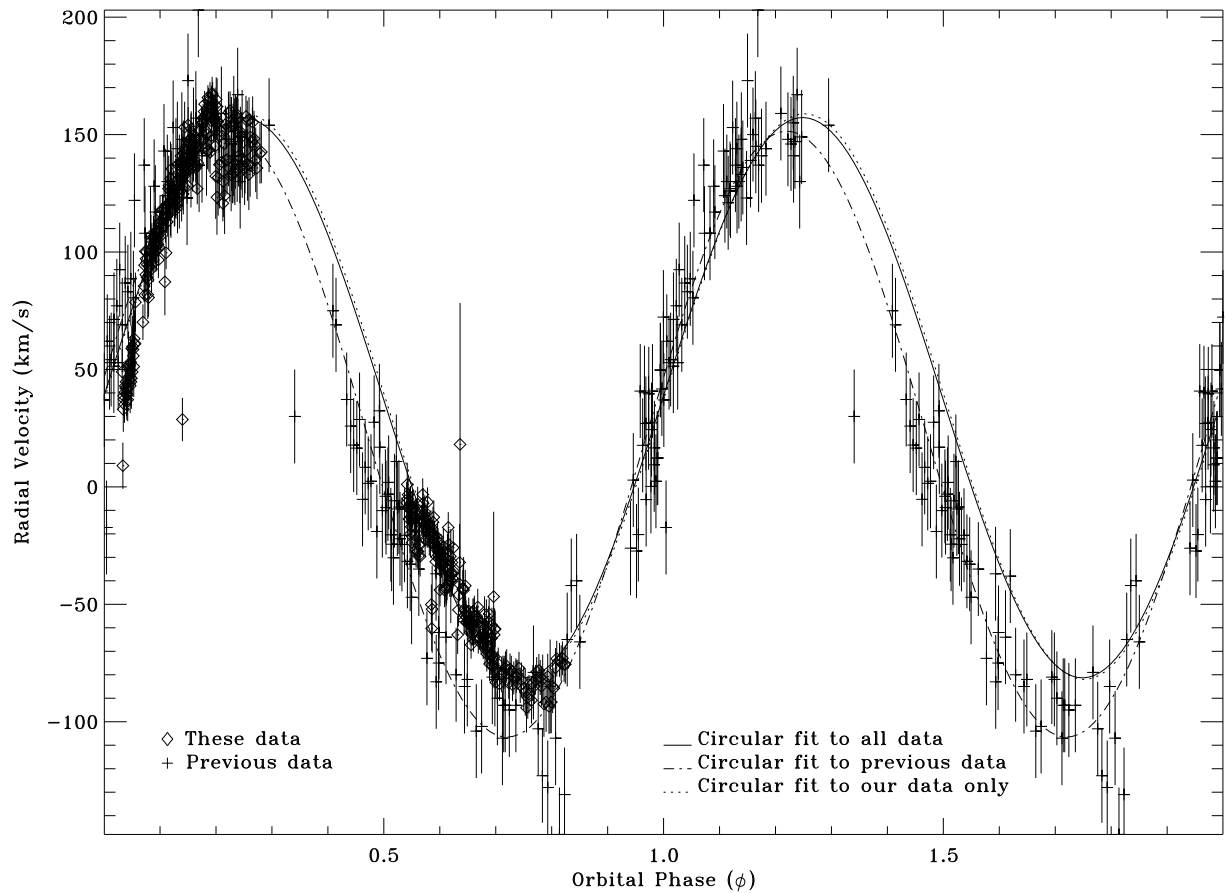}}
\noindent
\end{picture}
\vspace{125mm}
\caption{Radial velocity curve obtained by cross-correlating
  the absorption features of the companion star for the quiescent data
  sets against suitable templates. The radial velocities measured by
  Kraft (1964), Crampton \etal (1986), and Reinsch (1994) are also
  plotted. The solid line is the circular fit to all the data.  The
  dashed line is the fit to the data previous to this study. The
  dotted line is a fit to the data presented in this paper. Note that
  although two orbits are plotted, our data is only plotted once for
  clarity.}
\label{donor:rv:quiescence:fig2}
\end{figure*}

Cross-correlation of the absorption lines of \gk\ and the broadened
template yielded the radial velocities of the secondary star in the
binary, relative to that of the K1{\sc iv} template used. The radial
velocities obtained were then shifted to account for the radial
velocity of HD197964; $-$6.5~km\,s$^{-1}$ (Evans 1979).

We combined these results with previously published radial velocities
(Kraft 1964, Crampton \etal\ 1986, and Reinsch 1994), all obtained
during quiescence and carried out a search for periodicities using the
Lomb-Scargle algorithm (Scargle 1982). Based on this period search we
obtained a new ephemeris for \gk:

\begin{equation}
T_{0}(HJD) = 2450022.3465(0.0006) + 1.9968(0.0008) E
\end{equation}

We plot the radial velocities folded using this ephemeris in
Fig.~\ref{donor:rv:quiescence:fig2}.  We did not include the radial
velocities measured during outburst because they show increases and
decreases steeper than those seen during quiescence (see MSR99 for a
comparison between radial velocities measured during quiescence and
outburst). Although we do not have an explanation for this behaviour,
we decided not to include the outburst data for consistency. The
errors on individual measurements previous to this study were assumed
to be equal to the mean error of 20~km\,s$^{-1}$ (Reinsch 1994).  We
fitted all the combined data with a sinusoid:

\begin{equation}
V = \gamma + K_{\rm K} \sin 2 \pi [\phi - \phi_0]
\end{equation}

and plot it with a solid line in Fig.~\ref{donor:rv:quiescence:fig2}.
The best fit provides $\gamma$= 37.9 $\pm$ 0.3 \kmsec, and K$_{\rm
  K}$= 119.3 $\pm$ 0.6 \kmsec.  For comparison, we fitted only the
radial velocities previous to this work with a circular function
obtaining $\gamma$= 22.5 $\pm$ 1.7, and K$_{\rm K}$= 129.0 $\pm$ 2.7
\kmsec.  The uncertainties of the fit parameters from the combined
data are smaller than those of the published data, and the two
solutions for K$_{\rm K}$ are consistent within 3-$\sigma$. We also
fitted a sinusoidal to our data only and obtained $\gamma$=
40.8$\pm$0.7 \kmsec\ and K$_{\rm K}$= 120.5 $\pm$ 0.7 \kmsec. This
value for K$_{\rm K}$ is consistent within 1-$\sigma$ with that
obtained from the combined data. For clarity we plot two orbits in
Fig.~\ref{donor:rv:quiescence:fig2}. Our data is plotted only once so
it is easier to see how good the fit is for previously published data.

We noticed that between orbital phases 0.4 -- 0.8 the previously
published data shows large deviations from the fit to all the data.
This is not surprising considering that the radial velocity
measurements were done using different radial velocity standards, of
different spectral types in some cases. We mentioned before that Kraft
(1964) had noticed spectral changes in the secondary during his
observations which could have been caused by irradiation of the
secondary. This would explain the deviations from the fit observed.
For this reason we have decided to adopt as the best solution the
fit to only our data, as it was obtained from a consistent dataset.

A second-order sinusoidal fit has been found in the past to provide a
statistically more-significant fit than a circular one. This results
from the irradiation of a fraction of the stellar surface (Friend
\etal\ 1990).  Such a fit in this case provides no significant
eccentricity, verifying we have not detected an irradiated atmosphere
in this data.

\subsection{Emission line velocities}
\label{emission}

\begin{figure}
\begin{picture}(100,0)(10,20)
\put(0,0){\includegraphics{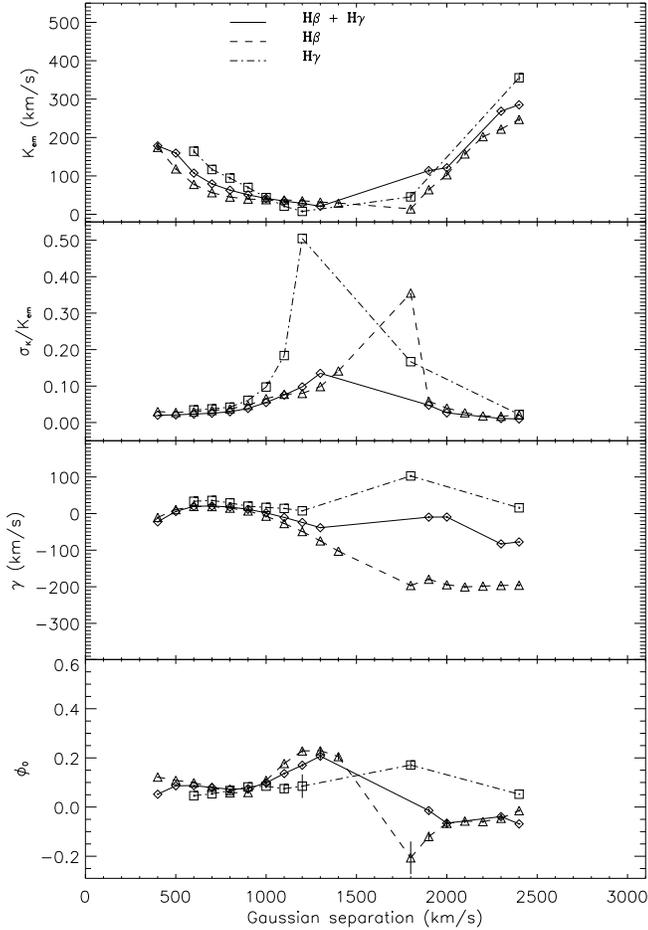}}
\noindent
\end{picture}
\vspace{130mm}
\caption{Diagnostic  diagram  of  the Balmer   lines, showing  the
  radial velocity semiamplitude K$_{\rm em}$ and its fractional error,
  $\sigma_{K}/{\rm K}_{\rm em}$, as well as the systemic velocity,
  $\gamma$, and the zero phase, $\phi_{0}$. The zero phase is relative
  to the ephemeris derived in Section~3.2. The dashed and dash-dotted
  lines represent the results obtained after applying the Schneider \&
  Young (1980) method to \hb\ and \hgam\ respectively whereas the
  solid line presents the results for the sum of both lines.}
\label{quiescence:wdrv:fig1}
\end{figure}

\begin{figure}
\begin{picture}(100,0)(10,20)
\put(0,0){\includegraphics{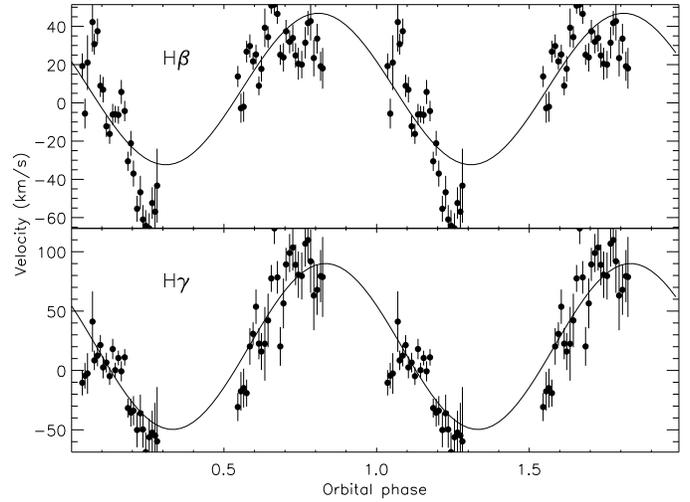}}
\noindent
\end{picture}
\vspace{70mm}
\caption{The radial velocity curves for H$\beta$ and H$\gamma$ from
  measurements with a double Gaussian separation of 900 km\,s$^{-1}$.}
\label{quiescence:wdrv:fig2}
\end{figure}

\begin{figure}
\begin{picture}(100,0)(10,20)
\put(0,0){\includegraphics{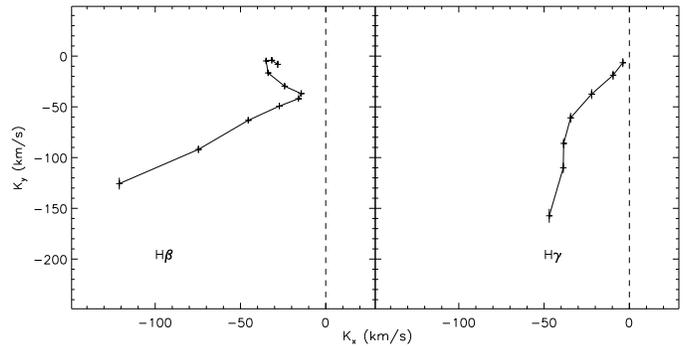}}
\noindent
\end{picture}
\vspace{50mm}
\caption{The light centres of the disc deduced from the radial
  velocity fits of \hb\ and \hgam. The position moves from left to
  right as the Gaussian separation increases which is equivalent to
  measuring the velocities closer to the white dwarf. For \hb\ the
  Gaussian separations go from 400 to 1400 \kmsec\ in 100 \kmsec\ 
  steps whereas for \hgam\ they go from 600 to 1200 \kmsec\ in
  100\kmsec\ steps. In \hb\ we have ignored separations above 900
  \kmsec\ in the extrapolation to the K$_y$ axis as they start
  departing from K$_x$=0.}
\label{quiescence:wdrv:fig3}
\end{figure}

Crampton \etal\ (1986) measured the radial velocity semi-amplitudes of
the wings of \hb\ to be $34 \pm 5$ \kmsec\ and combined with their
absorption line measurements reported a mass ratio $q$ ($M_{\rm
  K}$/$M_{\rm WD}$) of 0.28, where $M_{\rm WD}$ and $M_{\rm K}$ are
the masses of the primary and secondary star respectively. Although
they used only \hb\ because the variations in the radial velocities
for the other emission lines showed more scatter, it is clear from
their Fig.~3 that even the scatter in \hb\ gives a very unreliable
fit. Adopting the emission line velocity to be the same as the white
dwarf requires the assumption that the disc emits axisymmetrically,
however there are many examples where this is not the case (Kaitchuck
\etal\ 1994).  Axisymmetry can generally be minimised by sampling the
velocities of the high velocity line wings only since the inner disk
is expected to be more axisymmetric than the outer (Marsh 1988) and we
apply this method to \gk.

We use the double-gaussian convolution technique of Schneider \& Young
(1980).  This convolves each continuum-subtracted,
logarithmically-rebinned spectrum with two Gaussians of identical
width, FWHM = 200~km\,s$^{-1}$, and of separations varying
systematically between 400 and 3000~km\,s$^{-1}$.  The resulting
radial velocities were fitted using:
\begin{equation}
V = \gamma - {\rm K}_{\rm em} \sin 2 \pi [\phi - \phi_0]
\end{equation}
and the parameters of the fit are plotted in the diagnostic diagram
(Shafter, Szkody \& Thorstensen 1986) shown in
Fig.~\ref{quiescence:wdrv:fig1}.  Faster gas velocities are sampled as
the Gaussian separation increases.  The value of K$_{\rm em}$ depends
sensitively on the segment of line that is sampled.  Commonly, K$_{\rm
  WD}$ is considered to coincide with K$_{\rm em}$ when K$_{\rm em}$
is stable over a range of Gaussian separations, i.e.  $36 \pm 13$
\kmsec\ in this case, or at the point just before $\sigma_{K}/{\rm
  K}_{\rm em}$ increases sharply, i.e. at a Gaussian separation of
$\sim$900\,\kmsec\ wich corresponds to $\sim$40 \kmsec\ in this
instance.  The actual radial velocities and the fits obtained are
plotted in Fig.~\ref{quiescence:wdrv:fig2} for measurements taken when
the Gaussian separation is 900\kmsec. Another approach to the
diagnostic diagram is the light centres method (Marsh 1988) in which
we plot the radial velocities measured in velocity space,
extrapolate the line of points obtained to the K$_y$ axis and read off
the value. Using this method we find K$_{\rm em}\sim 30$\,\kmsec\ for
\hb\ and 0\,\kmsec\ for \hgam\ (see Fig.~\ref{quiescence:wdrv:fig3}).

However, both approaches are not particularly justified and we find it
impossible to determine K$_{\rm WD}$ from these measurements.

\subsection{Rotational line broadening}

Fortunately, if we make the assumption that the secondary stars
rotation is tidally-locked to the orbital period we can determine the
mass ratio of the binary using only the projected radial velocity
semi-amplitude and rotational velocity of the secondary
(Section~\ref{donor:vsini:q}).  This rotational quantity can be
measured from the width of the absorption lines.  Calculation of the
masses of the components of binary systems from V$_{\rm K}\sin i$
measurements is considered a more accurate method than measuring the
emission line velocities, although Bleach \etal\ (2000) provide some
caveats.

We used the radial velocity fit obtained in Section~3.2 to subtract
the orbital velocity of the absorption features in \gk, and averaged
the continuum subtracted spectra for each night.  We convolved the
K1{\sc iv} template spectrum with a broadening function in which we
include a linear limb-darkening coefficient of 0.53 (Gray 1992; van
Hamme, 1993), obtaining a series of templates corresponding to V$_{\rm
  K}\sin i$ = 5--150~km\,s$^{-1}$.  Then we applied the same optimal
subtraction technique as in Section~3.1 to find which broadened
templates best matched the lines in the average spectrum. This
technique has been applied by many authors to close binary systems
(e.g.  Southwell \etal, 1995; Marsh \etal, 1994; Drew, Jones \& Woods,
1993; Friend \etal, 1990).

The INT  and McDonald spectra yielded V$_{\rm  K}\sin$ i =  64.3 $\pm$
11.0~km\,s$^{-1}$, and 58.7  $\pm$  14.2~km\,s$^{-1}$ respectively.   We
adopt V$_{\rm K}\sin i$ = 61.5 $\pm$ 11.8~km\,s$^{-1}$, the mean of the
values found for the quiescent data.

\subsection{Stellar masses}
\label{donor:vsini:q}

\begin{figure*}
 \begin{picture}(100,0)(10,20)
\put(0,0){\includegraphics{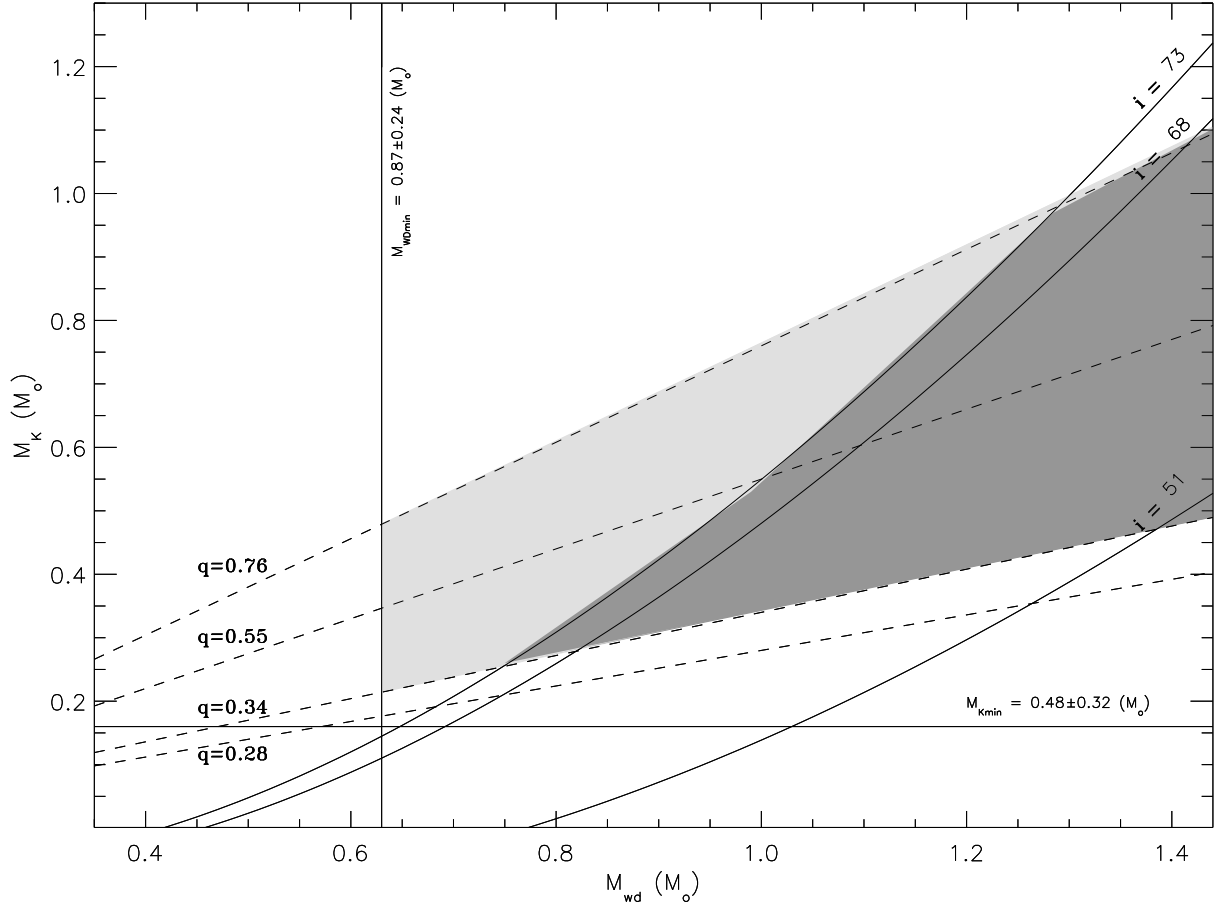}}
\noindent
\end{picture}
\vspace{125mm}
\caption{Possible values for M$_{\rm  K}$ and M$_{\rm WD}$ from the
  calculated mass function. Since the system does not show eclipses,
  $i$ must be smaller than 73$^{\circ}$ (limit obtained from the
  maximum value of $q$ we calculate). The minimum value possible for
  M$_{\rm K}$ found from this work is 0.48 $\pm$ 0.32 M$_\odot$.  This
  implies a minimum value for M$_{\rm WD}$ of 0.87$\pm$0.24 M$_\odot$.
  The maximum possible value for M$_{\rm WD}$ is the Chandrasekhar
  limit (1.44 M$_\odot$). The light grey-shaded region gives the
  possible mass solutions for \gk. The limits of this region are the
  minimum and maximum possible value of M$_{\rm WD}$ and the minimum
  and maximum values of $q$ found from this work . The dark
  grey-shaded region includes the limit imposed by the inclination of
  the system.  Several mass ratio curves are also plotted.}
\label{donor:vsini:q:fig1}
\end{figure*}

From K$_{\rm K}$= 120.5 $\pm$ 0.7\,\kmsec, measured in Section~3.2, we
can obtain a value for the mass function of white dwarf of 0.362 $\pm$
0.006 M$_\odot$ using:
\begin{equation}
\frac{M_{\rm WD} \sin^{3} i}{(1 + q)^2}=
\frac{K_{\rm K}^{3}  P}{2\pi G} \quad
\label{data:anal:rv:eq4}
\end{equation}

In Section~\ref{emission} we defined the stellar mass ratio $q$, where
\begin{equation}
q = \frac{M_{\rm K}}{M_{\rm WD}} = 
\frac{K_{\rm WD}}{K_{\rm K}}  \quad
\label{data:anal:rv:eq3}
\end{equation}
The rotational, $V_{\rm K}\sin i$, and radial velocity semi-amplitude,
$K_{\rm K}$, of the secondary star are related by
\begin{equation}
\label{donor:vsini:q:eq1}
V_{\rm K}\sin i = K_{\rm K} (1+q) R_{\rm L_{K}}(q)
\end{equation}
where $R_{\rm L_{K}}(q)$ is the spherically-averaged Roche lobe radius
of the   companion star, scaled to  the   stellar separation. Eggleton
(1983)  has supplied an  analytical approximation for this, correct to
$\sim$ 1\%:
\begin{equation}
\label{donor:vsini:q:eq2}
R_{\rm L_{K}}(q) = \frac{0.49 q^{2/3}}{0.6 q^{2/3} + \ln(1 + q^{1/3})} ,\quad
0 < q < \infty
\end{equation}
Consequently  $q = 0.55 \pm 0.21$.  This value does not  agree with $q =
0.28$ measured by combining absorption and emission line velocities by
Crampton \etal\ (1986) and the implied velocity semi-amplitude for the 
white dwarf is K$_{\rm WD}$= 66.3 $\pm$ 25.3 \kmsec.

Substituting $q$ into Eq.~4, we obtain a lower limit for the mass of
the white dwarf of M$_{\rm WD}$ = 0.87 $\pm$ 0.24M$_\odot$.  Combining
this value with $q$ we then have a lower limit on M$_{\rm K}$ of 0.48
$\pm$ 0.32 M$_\odot$.

Reinsch (1994) observed \gk\ at orbital phases during which potential
eclipses would happen but did not measure any eclipse effects in the
line wings. This implies that the white dwarf and inner accretion
region are not eclipsed. Eclipses of the compact object and inner
accretion regions require an orbital inclination $i > \cos^{-1} \rm
R_{\rm L_{\rm K}}$, where R$_{\rm L_{\rm K}}$ is the
spherically-averaged Roche lobe radius of the secondary star, scaled
to the stellar separation, defined in Eq.~6.

Assuming a value for $q$ of 0.55 $\pm$ 0.21, we obtain that eclipses
would occur if $i$ was greater than 73$^{\circ}$. Fig.~7 represents
the possible values for the masses of the components of the system
depending on the mass ratio $q$ and the inclination $i$ of the binary.
Solutions are presented between the mass function value of
0.362M$_{\odot}$, and the Chandrasekhar white dwarf mass M$_{\rm WD}$
=~1.44M$_{\odot}$. The grey-shaded regions give the possible solutions
to the mass of the system (see figure caption for more details).

\section{Conclusion}

We have measured the mass ratio of \gk, $q = 0.55 \pm 0.21$ using the
width and orbital velocity of absorption lines from the secondary star,
and determine lower limits for the stellar masses, M$_{\rm WD} \geq$
0.87\,$\pm$\,0.24 and M$_{\rm K}\geq$ 0.48\,$\pm$\, 0.32\,M$_{\odot}$,
without using dubious measurements of K$_{\rm WD}$.  The spectral type
of the secondary star remains constant over the orbital cycle
indicating that the inner face of the companion is not significantly
irradiated by accretion radiation.

\section*{ACKNOWLEDGEMENTS}

MDS was supported by PPARC grant K46019.  PDR acknowledges the support
of the Nuffield Foundation via a grant to newly qualified lecturers in
science to assist collaborative research.  The reduction and analysis
of the data were carried out on the Sussex node of the STARLINK
network. We thank Tom Marsh for providing his reduction software. The
Isaac Newton Telescope is operated on the island of La Palma by the
Isaac Newton Group in the Spanish Observatorio del Roque de los
Muchachos of the Instituto de Astrof\'{\i}sica de Canarias.

\label{lastpage}


\begin{thebibliography}{99}
  

\bibitem{55} Bleach J.\,N., Wood J.\,H., Catal\'{a}n M.\,S., Welsh
  W.\,F., Robinson E.\,L., Skidmore W., 2000, MNRAS, 312, 70

\bibitem{18} Campbell L., 1903, Harv. Ann., 48, 90

\bibitem{2}  Crampton D., Cowley A.\,P., Fisher W.\,A., 1986, ApJ,
  300, 788

\bibitem{402} Dhillon V.\,S., Marsh T.\,R., Jones D.\,H.\,P., 1991,
  MNRAS, 252, 342

\bibitem{103}Drew J.\,E., Jones D.\,H.\,P., Woods J.\,A., 1993, MNRAS,
  260, 803
\bibitem{104}Eggleton P.\,P., 1983, ApJ, 268, 368
\bibitem{44} Evans D.\,S., 1979, IAU Symp., 30, 57 
\bibitem{105}Friend M.\,T., Martin J.\,S., Smith R.\,C., Jones
  D.\,H.\,P., 1990, MNRAS, 246, 637
\bibitem{22} Gallagher J.\,S., Oinas V., 1974, PASP, 86, 952 
\bibitem{995} Georgelin Y., 1967, Publ. Obs. Haute-Provence, 9, 27
\bibitem{300}Gray D.\,F., 1992, The Observation and Analysis of
  Stellar Photospheres. Cambridge University Press, Cambridge
\bibitem{994} Griffin R.\,F., Redman R.\,O., 1960, MNRAS, 120, 287
\bibitem{993} Hoffleit D., Warren Jr W.\,H., 1991, Bright Star
  Catalogue 5th Revised Ed.
\bibitem{29} Horne K., 1986, PASP, 98, 609 
\bibitem{999} Houk N., Smith-Moore M., 1988, Michigan Soectral Survey,
  Ann Arbor, Dept. Astron, Univ. Michigan
  
\bibitem{401} Kaitchuck R.\, H., Schlegel E.\,M., Honeycutt R.\,K.,
  Horne K.,. Marsh T.\,R., White II J.\,C., Mansperger C.\,S., 1994,
  ApJSS, 93, 519-530

\bibitem{23} Kraft R.\,P., 1964, ApJ, 139, 457

\bibitem{404} Marsh T.\,R., 1988, MNRAS, 231, 1117

\bibitem{24} Marsh T.\,R., Robinson, E.\,L., Wood, J.\,H., 1994,
  MNRAS, 266, 137 
\bibitem{9}  Mattei J.\,A., Bortle, J., Dillon, W., Royer, R.,
  Schmeer, P., Komorous, R.\,A., Osorio, J.\,R., McKenna, J., 1996,
  IAU Circ.\ 6325 
  1985, A\&A, 149, 470

\bibitem{3}  Morales-Rueda L., Still M.\,D., Roche P., 1996, MNRAS, 283, L58
\bibitem{62} Morales-Rueda L., Still M.\,D., Roche P., 1999, MNRAS, 306, 753

\bibitem{998} Nassau J.\,J., van Albada G.\,B., 1947, ApJ, 106, 20
\bibitem{201} Oke J.\,B., Gunn J.\,E., 1983, ApJ, 266, 713

\bibitem{980} Patterson J., 1984, ApJS, 54, 443
\bibitem{11} Reinsch K., 1994, A\&A, 281, 108
\bibitem{996} Roman N.\,G., 1952, ApJ, 116, 122
\bibitem{15} Sabbadin F., Bianchini A., 1983, A\&AS, 54, 393 
\bibitem{356} Scargle J.\,D., 1982, ApJ, 263, 835
\bibitem{203} Schneider D.\,P., Young P.\,J., 1980, ApJ, 238, 946
\bibitem{205} Shafter A.\,W., Szkody P., Thorstensen J.\,R., 1986,
  ApJ, 308, 765
\bibitem{60} Skidmore W., Welsh W.\,F., Wood J.\,H., Stiening R.\,F.,
  1997, MNRAS, 288, 189
\bibitem{200}Smith D.\,A., Dhillon V.\,S., 1998, MNRAS, 301, 767
\bibitem{108}Southwell K.\,A., Still M.\,D., Smith R.\,C., Martin
  J.\,S., 1995, A\&A, 302, 90
\bibitem{202}Stone R.\,P.\,S., 1977, ApJ, 218, 767
\bibitem{32} Tonry J., Davis, M., 1979, AJ, 84, 1511
\bibitem{107}van Hamme W., 1993, AJ, 106, 2096 

\bibitem{997} Yoss K.\,M., 1961, ApJ, 134, 809
\bibitem{61} Zhang E., Robinson E.\,L., Ramseyer T.\,F., Shetrone
  M.\,D., Stiening R.\,F., 1991, ApJ, 381, 534

\end{thebibliography}
\end{document}